\documentclass[aps,pra,onecolumn,showpacs]{revtex4}

\usepackage{graphics}
\usepackage{epsfig}

\begin{document}

\draft

\title{Information transmission through lossy bosonic memory channels}

\author{Giovanna Ruggeri$^{1,2}$, Giulio Soliani$^{1,2}$, 
Vittorio Giovannetti$^{3}$ and Stefano Mancini$^{4,5}$}

\affiliation{
$^{1}$Dipartimento di Fisica dell'Universit\`a di Lecce, I-73100 Lecce, Italy\\
$^{2}$INFN Sezione di Lecce, I-73100 Lecce, Italy\\
$^{3}$NEST-INFM \& Scuola Normale Superiore, I-56126 Pisa, Italy\\
$^{4}$Dipartimento di Fisica dell'Universit$\grave{a}$ di Camerino,
I-62032 Camerino, Italy\\
$^{5}$INFN Sezione di Perugia, I-06123 Perugia, Italy}


\date{today}

\begin{abstract}
We study the information transmission through a quantum channel, defined over a continuous alphabet and losing its energy en route, in presence of correlated noise among different channel uses.
We then show that entangled inputs improve the rate of transmission of such a channel.
\end{abstract}

\maketitle

\section{Introduction}

Communication of information requires an encoding of the information into a physical system.  The laws of physics therefore govern the limits on processing and communication of information.  By modelling real world noise in terms of simpler models, the maximum rate for information transfer may be obtained.  
\cite{gall}.
Within the quantum framework this approach
has primarily been developed for memoryless channels \cite{nich}, where the noise on each transmitted state is treated as independent.  However, many real world communication channels experience noise which is modelled better by errors that are correlated between separate channel uses.  
As a consequence, recently, there has been an increasing interest
for quantum channels with memory effects 
 \cite{macc,altri,bm,bdb}, but only involving discrete alphabets (qubits).

Quantum communication with continuous alphabet provides an interesting alternative to 
the traditional discrete alphabet based approach~\cite{bp01}. 
A lot of efforts have been devoted to characterize continuous alphabet quantum 
channels  \cite{benn}. 
Among them we can mention the lossy bosonic channel, which consists of a collection of bosonic 
modes that lose energy en route from the transmitter to the receiver.
The use of entangled inputs 
among different channel uses 
has found to unaffected the classical capacity of 
such communication lines \cite{giov},
in close analogy to what happen for a wide class of qubit alphabet channels 
\cite{kr01}.
However, for the latter it was argued that entangled inputs may enhance the information
 transmission in presence of correlated noise (memory) \cite{macc}. 

Here, we study the information transmission through a lossy bosonic channel  in presence of memory effects. The latter are modeled by considering correlations among environments acting on different channel uses \cite{giovman}.
We then show that entangled inputs improve the rate of transmission of such a channel.

\section{The model}

A quantum channel that uses continuous alphabet can 
be modeled by a bosonic field mode
whose phase space quadratures enable for continuous 
variable encoding/decoding.
On $n$ uses of such a channel we have to consider $n$ independent bosonic modes,
described by annihilation 
operators ${\hat a}_k$ for $k=1,\cdots,n$.
Each ${\hat a}_k$ interacts with
an environment mode ${\hat b}_k$ through a 
beam splitter of transmittivity $\eta\in[0,1]$, thus modeling losses. 
The signal-noise coupling is then characterized by the transformation \cite{walls94}
\begin{eqnarray}
{\hat a}_k &\longrightarrow&\sqrt{\eta}\; {\hat a}_k - \sqrt{1-\eta} \; {\hat b}_k\,,
\nonumber\\
{\hat b}_k &\longrightarrow&\sqrt{\eta}\; {\hat b}_k + \sqrt{1-\eta} \; {\hat a}_k\,.
\label{unouno}
\end{eqnarray}
Let be ${\hat\rho}_{in}$ the density operator in the Hilbert space of the $n$ modes ${\hat a}_k$ representing the channel input.
The encoding  on $n$ uses of the channel is considered as a mixture of entangled  coherent states, 
defined as
 \begin{equation}
 |\psi(\mbox{\boldmath$\mu$})\rangle={\hat S}(r)\left [{\hat D}_{a_n} (\mu_n)|0 \rangle_{a_n} \ldots\ {\hat D}_{a_1}(\mu_1)|0\rangle_{a_1}\right],
 \label{psi}
  \end{equation}
where ${\hat D}_k (\mu_k)$ is the single $k$-mode displacement operator, corresponding to the complex number $\mu_k$ and
${\hat S}(r)$ denotes the $n$-mode squeeze operator, with $r$ the entanglement parameter between different channel uses ($r=0$ refers to absence of entanglement) \cite{walls94}.
We assume the states  $|\psi(\mbox{\boldmath$\mu$})\rangle$ weighted with the Gaussian probability distribution
   \begin{equation}
P(\mbox{\boldmath$\mu$})=\prod^n_{k=1}P_k(\mu_k)\,,\qquad
P_{k}(\mu_k)\equiv\frac{1}{\pi N_k} e^{-|\mu_k|^2/N_k}\,,
\label{Pimu}
 \end{equation}
where $N_k$ is the average photon number per channel use.
For the sake of simplicity, we restrict our analysis to the case $N_k=N,\; \forall k$.
However, by virtue of entanglement, the effective average photon number per channel use will be 
$N_{eff}=N+\sinh^2 r$. Then, we can write
 \begin{equation}
 \rho_{in}=\int\,d\mbox{\boldmath$\mu$} P(\mbox{\boldmath$\mu$}) \rho_{in}(\mbox{\boldmath$\mu$})\,,
 \qquad  \rho_{in}(\mbox{\boldmath$\mu$})\equiv|\psi(\mbox{\boldmath$\mu$})\rangle \langle\psi(\mbox{\boldmath$\mu$})|.
 \label{rhoin}
 \end{equation}
The  Wigner function corresponding to $\rho_{in}(\mbox{\boldmath$\mu$})$ is \cite{bp01}
 \begin{eqnarray} 
 W_{in}(\mbox{\boldmath$u$};\mbox{\boldmath$\mu$})=\left(\frac{2}{\pi}\right)^n
 \exp\left[-\mbox{\boldmath$u$}{\cal A}^{\prime}_r \mbox{\boldmath$u$}^T-\mbox{\boldmath$\mu$}{\cal A}^{\prime}_r \mbox{\boldmath$\mu$}^T+2\mbox{\boldmath$\mu$}{\cal A}^{\prime}_r \mbox{\boldmath$u$}^T\right]\,,
 \label{Win}
 \end{eqnarray}
where we now intend $\mbox{\boldmath$u$}$, $\mbox{\boldmath$\mu$}$ as vectors in ${\bf R}^{2n}$ 
 \begin{eqnarray} 
\mbox{\boldmath$u$}&\equiv&\left(
 x_1,\ldots,x_n,p_1,\ldots,p_n\right)\,,\\
 \mbox{\boldmath$\mu$}&\equiv&\left(
 \mu_1^R,\ldots,\mu_n^R,\mu_1^I,\ldots,\mu_n^I\right)\,,
 \end{eqnarray}
with $x_k$, $p_k$ quadrature variables of ${\hat a}_k$
and $\mu^R_k$, $\mu^I_k$ real and imaginary part of  $\mu_k$
($T$ denotes the transpose operation).
Moreover, ${\cal A}^{\prime}_r$ is a real $2n\times 2n$ matrix of the form\footnote{Throughout the paper we denote respectively $4n\times 4n$, $2n\times 2n$, $n\times n$ matrices by simple, primed, double primed, calligraphic letters.}
 \begin{equation}\label{Ar}
 {\cal A}^{\prime}_r=\frac{2}{n}
 \left(
 \begin{array}{cc}
 {\cal A}^{\prime\prime}_r  &  0  \\
 0  & {\cal A}^{\prime\prime}_{\,-r}
 \end{array}
 \right)\,,
 \end{equation}
with the $n\times n$ matrix
 \begin{equation}
  {\cal A}^{\prime\prime}_r=
 \left(
 \begin{array}{cccc}
 e^{-2r}+(n-1) e^{2r}  & e^{-2r}-e^{2r} & \ldots &   e^{-2r}-e^{2r}  \\
 e^{-2r}-e^{2r} & e^{-2r}+(n-1) e^{2r}  & \ldots &  e^{-2r}-e^{2r} \\
 \vdots &   &  \ddots & \vdots  \\
  e^{-2r}-e^{2r} &   &   &  e^{-2r}+(n-1) e^{2r}
 \end{array}
 \right)\,.
 \end{equation}
 For a memoryless channel the environment acts independently on each
${\hat a}_k$. This can be modeled by assuming the modes ${\hat b}_k$ to be in the same state, e.g. 
the vacuum.
A memory channel is characterized by non-trivial correlations between the environment actions on the different channel uses. According to \cite{giovman}, this can be accounted for by an $n$-mode squeezed vacuum state, whose Wigner function reads \cite{bp01}
 \begin{eqnarray}
W_m(\mbox{\boldmath$v$})=\left(\frac{2}{\pi}\right)^n
\exp\left[-\mbox{\boldmath$v$}{\cal A}^{\prime}_s \mbox{\boldmath$v$}^T\right]\,,
\end{eqnarray}
where $\mbox{\boldmath$v$}$ is vector in ${\bf R}^{2n}$ 
 \begin{eqnarray} 
 \mbox{\boldmath$v$}&\equiv&\left(
 y_1,\ldots,y_n,q_1,\ldots,q_n\right)\,,
 \end{eqnarray}
 with $y_k$, $q_k$ quadrature variables of ${\hat b}_k$.
Moreover, $s$ is the entanglement (memory) parameter describing the correlated noise effects ($s=0$ corresponds to memoryless case), and ${\cal A}^{\prime}_s$ is given by Eq.(\ref{Ar}) with $r\to s$.

To go further on, let us define the vectors in ${\bf R}^{4n}$ 
\begin{eqnarray}
\mbox{\boldmath$\gamma$}\equiv(\mbox{\boldmath$u$},\mbox{\boldmath$v$})\,,\qquad
\mbox{\boldmath$\theta$}\equiv(\mbox{\boldmath$u$},\mbox{\boldmath$0$})\,,\qquad
\mbox{\boldmath$\kappa$}\equiv(\mbox{\boldmath$\mu$},\mbox{\boldmath$0$})\,.
\end{eqnarray}
Then, we can write the total (input plus memory) Wigner function as
\begin{equation}\label{Wtot}
W_{tot}\left(\mbox{\boldmath$\gamma$};\mbox{\boldmath$\kappa$}\right)=W_{in}\left(\mbox{\boldmath$u$};\mbox{\boldmath$\mu$}\right)W_m\left(\mbox{\boldmath$v$}\right)=\left(\frac{2}{\pi}\right)^{2n}
\exp\left[
-\mbox{\boldmath$\gamma$}{\cal A}\mbox{\boldmath$\gamma$}^T+2\mbox{\boldmath$\kappa$}{\cal A}\mbox{\boldmath$\gamma$}^T-\mbox{\boldmath$\kappa$}{\cal A}\mbox{\boldmath$\kappa$}^T\right]\,,
\end{equation}
where ${\cal A}$ is a real $4n \times 4n$ matrix of the form
\begin{equation}
{\cal A}=
\left(
\begin{array}{cc}
{\cal A}^{\prime}_r  &  0  \\
0  &  {\cal A}^{\prime}_s   
\end{array}
\right)\,.
\end{equation}
As a consequence of Eq.(\ref{unouno}), the signal-noise coupling corresponds to the change of variables 
\begin{equation}\label{bs}
\mbox{\boldmath$\gamma$}^T\longrightarrow {\cal B}\mbox{\boldmath$\gamma$}^T\,
\end{equation}
produced by the unitary beam splitter matrix
\begin{equation}\label{O}
{\cal B}=
\left(
\begin{array}{ccc}
{\cal B}^{\prime}_1  &   {\cal B}^{\prime}_2   \\
 -{\cal B}^{\prime}_2 &  {\cal B}^{\prime}_1 
\end{array}
\right)\,,
\end{equation}
with the $2n\times 2n$ matrices
${\cal B}^{\prime}_1={\rm diag}\; \sqrt{\eta}$ and ${\cal B}^{\prime}_2={\rm diag}\; \sqrt{1-\eta}$.
Then, by using (\ref{bs}) in (\ref{Wtot}) and integrating over the memory variables $v$,  we get the output Wigner function 
\begin{equation}
W_{\rm out}(\mbox{\boldmath$\theta$}; \mbox{\boldmath$\kappa$})=\left(\frac{2}{\pi}\right)^{2n}
\exp\left[
-\mbox{\boldmath$\theta$}\,{\cal G}\,\mbox{\boldmath$\theta$}^T+2\mbox{\boldmath$\kappa$}\,{\cal F}\,\mbox{\boldmath$\theta$}^T-\mbox{\boldmath$\kappa$}\,{\cal A}\,\mbox{\boldmath$\kappa$}^T
\right]\,,
\end{equation}
where 
\begin{equation}
{\cal F}={\cal A}{\cal B}\,,\qquad
{\cal G}={\cal B}^T{\cal A}{\cal B}\,. 
\end{equation}
Finally, we consider decoding at the output by means of a heterodyne measurement,
that is by projecting the output density matrix on an $n$-mode coherent state $|\mbox{\boldmath$\zeta$}\rangle$.
Now, the matrix element  $\langle \mbox{\boldmath$\zeta$}|\rho_{out}|\mbox{\boldmath$\zeta$}\rangle$ can be interpreted as the conditional probability, explicitly given by
\begin{equation}\label{Pzm}
P\left(\mbox{\boldmath$\zeta$}|\mbox{\boldmath$\mu$}\right)=\left(\frac{2}{\pi}\right)^{n}
\int\,d\mbox{\boldmath$\theta$}\, W_{out} \left(\mbox{\boldmath$\theta$};\mbox{\boldmath$\kappa$}\right)\exp\left[-(\mbox{\boldmath$\xi$}-\mbox{\boldmath$\theta$}){\cal L} (\mbox{\boldmath$\xi$}-\mbox{\boldmath$\theta$})^{T}\right]\,.
\end{equation}
Here $\mbox{\boldmath$\xi$}$ should be intended as the $4n$ real component vector of the type 
\begin{equation}
\mbox{\boldmath$\xi$}\equiv\left(\mbox{\boldmath$\zeta$},\mbox{\boldmath$0$}\right),\qquad
\mbox{\boldmath$\zeta$}\equiv\left(\zeta_1^R,\ldots,\zeta_n^R,
\zeta_1^I,\ldots,\zeta_n^I\right)\,,
\end{equation}
and ${\cal L}$ as the $4n \times 4n$ matrix  
\begin{equation}
{\cal L}=2
\left(
\begin{array}{cc}
 {\cal I}^{\prime} &   0 \\
 0   &   0   
\end{array}
\right)\,,
\end{equation}
with ${\cal I}^{\prime}$ the $2n\times 2n$ identity matrix.

By making use of Gaussian integrals,
the conditional probability (\ref{Pzm}) can be rewritten
\begin{equation}\label{Pzmexp}
P\left(\mbox{\boldmath$\zeta$}|\mbox{\boldmath$\mu$} \right)=\frac{2^{3n}}{\pi^n \sqrt{\det \left({\cal G}+{\cal L}\right)}} \exp\left[-\mbox{\boldmath$\mu$}  {\cal R}^{\prime} \mbox{\boldmath$\mu$}^T+ \mbox{\boldmath$\zeta$} {\cal S}^{\prime} \mbox{\boldmath$\mu$}^T -\mbox{\boldmath$\zeta$} {\cal T}^{\prime} \mbox{\boldmath$\zeta$}^T\right]\,,
\end{equation}
where the  $2n \times 2n$ matrices ${\cal R}^{\prime}$, ${\cal S}^{\prime}$, ${\cal T}^{\prime}$ are submatrices, whose elements belong to the first $2n$ rows and columns of  the matrices
\begin{eqnarray}
{\cal R}&=&{\cal A} -{\cal F} \left( {\cal G}+{\cal L}\right)^{-1} {{\cal F}}^T \,,\\
{\cal S}&=& 2 {\cal L}  \left( {\cal G}+{\cal L}\right)^{-1} {{\cal F}}^T \,,\\
{\cal T}&=& {\cal L} -{\cal L}  \left( {\cal G}+{\cal L}\right)^{-1} {\cal L}\,.
\end{eqnarray}
Eq.(\ref{Pzmexp}) allows us to also derive the joint probability
\begin{eqnarray}
P\left(\mbox{\boldmath$\zeta$},\mbox{\boldmath$\mu$} \right)&\equiv& P\left(\mbox{\boldmath$\zeta$}|\mbox{\boldmath$\mu$}\right)P\left(\mbox{\boldmath$\mu$} \right)\nonumber\\
&=&\frac{2^{3n}}{\pi^{2n} N^n \sqrt{\det\left({\cal G}+{\cal L}\right)}} \exp\left[-\mbox{\boldmath$\mu$} \left({\cal R}^{\prime}+{\cal I}^{\prime}/N\right) \mbox{\boldmath$\mu$}^T+\mbox{\boldmath$\zeta$} {\cal S}^{\prime} \mbox{\boldmath$\mu$}^T-\mbox{\boldmath$\zeta$} {\cal T}^{\prime} \mbox{\boldmath$\zeta$}^T\right]\,,
\end{eqnarray}
and the output probability 
\begin{eqnarray}
P\left(\mbox{\boldmath$\zeta$} \right)&\equiv&\int\,d\mbox{\boldmath$\mu$} P\left(\mbox{\boldmath$\zeta$},\mbox{\boldmath$\mu$}\right)\nonumber\\
&=&\frac{2^{3n}}{\pi^n N^n \sqrt{\det\left({\cal G}+{\cal L}\right)} \sqrt{\det\left({\cal R}^{\prime}+{\cal I}^{\prime}/N\right)}} \exp\left[-\mbox{\boldmath$\zeta$} {\cal U}^{\prime} \mbox{\boldmath$\zeta$}^T\right]\,,
\end{eqnarray}
with the $2n \times 2n$ real matrix 
\begin{equation}
{\cal U}^{\prime}= {\cal T}^{\prime}-\frac{1}{4} {\cal S}^{\prime} \left({\cal R}^{\prime} + {\cal I}^{\prime}/N\right)^{-1} {{\cal S}^{\prime}}^T.
\end{equation}

From the definition of the Shannon entropy for a real continuous stochastic variable $\phi$
\begin{equation}
I\left(\phi\right)=- \int\,d\phi  P\left(\phi\right)\log_2{ P\left(\phi \right)}\,,
\label{Idef}
\end{equation}
we obtain
\begin{equation}
I(\mbox{\boldmath$\mu$})\,=\,\frac{1}{\ln2} \left[ n + \ln(\pi^n N^n)\right]\,,
\label{Imu}
\end{equation}

\begin{eqnarray}
&&I(\mbox{\boldmath$\zeta$})=\frac{2^{3n}}{N^n \ln2 \sqrt{\det\left({\cal G}+{\cal L}\right)} \sqrt{\det\left({\cal R}^{\prime}+{\cal I}^{\prime}/N\right)} \sqrt{\det {\cal U}^{\prime}}}\nonumber\\ 
&&\hspace{1 cm}\times\left[n-\ln \left( \frac{2^{3n}}{\pi^n N^n \sqrt{\det \left({\cal G}+{\cal L}\right)} \sqrt{\det\left({\cal R}^{\prime}+{\cal I}^{\prime}/N\right)}}\right) \right]\,,
\label{Ize}
\end{eqnarray}
and
\begin{equation}
I(\mbox{\boldmath$\zeta$}, \mbox{\boldmath$\mu$})\,=\,\frac{2^{3n}}{N^n \ln2 \sqrt{\det \left({\cal G}+{\cal L}\right)} \sqrt{\det {\cal V}}} \left[2n - \ln\left(\frac{2^{3n}}{\pi^{2n} N^n \sqrt{\det \left({\cal G}+{\cal L}\right)}}\right)\right]\,,
\label{Izemu}
\end{equation}
with the $4n \times 4n$ matrix
\begin{equation}
{\cal V} =
\left( 
\begin{array}{llcr}
{\cal R}^{\prime}+{\cal I}^{\prime}/N& -{\cal S}^{\prime}/2\\
 -{\cal S}^{\prime}/2 &   {\cal T}^{\prime}
\end{array}
\right).
\end{equation}

Now we can easily arrive at the mutual information, that is the information shared by the transmitter and the receiver when using the above described procedure
\begin{equation}
I_r\equiv I\left( \mbox{\boldmath$\zeta$}: \mbox{\boldmath$\mu$} \right)= I\left(\mbox{\boldmath$\mu$} \right)+I\left(\mbox{\boldmath$\zeta$} \right) -I\left(\mbox{\boldmath$\zeta$},\mbox{\boldmath$\mu$} \right)
\end{equation}
The mutual information over the number of uses $I_r/n$ defines the rate of transmission.
A useful quantity in our case would be the relative rate gain
\begin{equation}
g=\frac{I_r-I_{r=0}}{I_{r=0}}\,,
\end{equation}
which characterize the usefulness of entangled inputs.

\section{Results}

Although we derived a formal expression for the mutual information, it is almost impossible to learn something by inspection from it. Thus, 
we numerically study  the quantity $g$ as a function of the entanglement parameter $r$.

Since we want to bound the effective average photon number per channel use,   
in practice we fix $N_{eff}$ as the effective input photon number and we consider 
$N$ varying as function of $r$ ($N=N_{eff}-\sinh^2 r$), limiting the range of $r$ to those values for which $N\ge 0$.

\begin{figure}
\centering
\includegraphics[width=2.5in]{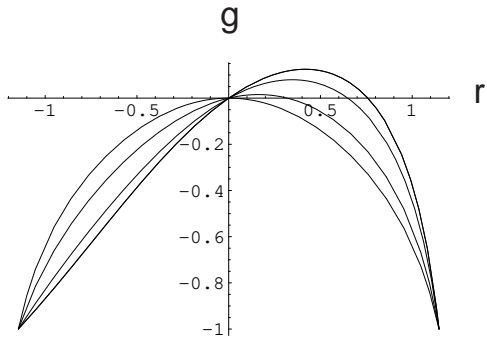}
\caption{Relative rate gain $g$ versus entanglement parameter $r$.
Curves from bottom to top are for $s=0$, $1$, $2$, $5$. 
The values of other parameters are $n=2$, $\eta=0.8$ and $N_{eff}=2$.}
\label{fig1}
\end{figure}

In Fig.\ref{fig1} the relative rate gain $g$ is shown versus the entanglement parameter $r$ for different values of the degree of memory $s$, for two uses of the channel and for an input photon number $N_{eff}=2$. 
For $s=0$ the behavior of $g$ is symmetric with respect to $r=0$, the value at which attains its maximum ($g=0$), that is, entangled inputs are no way useful in the memoryless case. 
As soon as the degree of memory increases the symmetry is broken and for
negative values of  $r$, the gain $g$ gets worse, while for positive value of $r$ gets better with respect to the $s=0$ case. In particular, for a limited range of positive $r$'s values it goes above $0$. This clearly shows an enhancement of the maximum transmission rate due to entangled inputs in case of memory channel with respect to memoryless one ($g=0$). 
Moreover, there is an optimal value of $r$ depending on the strength of the memory effect, for which $g$ attains its maximum. The latter is more pronounced for strong memory effect.

\begin{figure}
\centering
\includegraphics[width=2.5in]{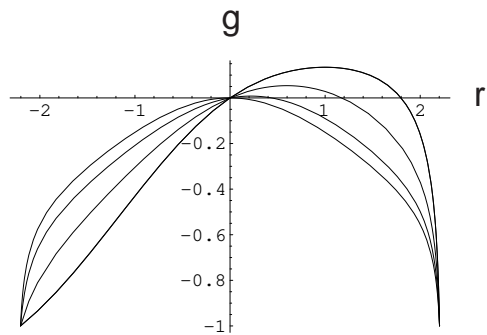}
\caption{Relative rate gain $g$ versus entanglement parameter $r$.
Curves from bottom to top are for $s=0$, $1$, $2$, $5$. 
The values of other parameters are $n=2$, $\eta=0.8$ and $N_{eff}=20$.}
\label{fig2}
\end{figure}

For an increased number of input photons the described effect is still well observable as shown by
Fig.\ref{fig2}. In this case the range of useful $r$ values is even larger.

Further investigations show that the quantities $I(\mbox{\boldmath$\mu$})$, $I(\mbox{\boldmath$\zeta$})$, $I(\mbox{\boldmath$\zeta$},\mbox{\boldmath$\mu$})$ are all linear in $n$. As matter of fact the coefficients in front of the square brackets in Eqs.(\ref{Imu}), (\ref{Ize}), (\ref{Izemu}) reduce to $1/\ln 2$ for all $n$ \footnote{We have analytically proved that for $n$ up to 4, while for higher values we only have numerical eveidences.}.
Thus, the results of Figs.\ref{fig1},\ref{fig2}  applies to any value of $n\ge2$.
This is a surprising peculiarity of the presented model because a possible enhancement of the transmission rate is usually ascribed to the non additivity the channel capacity. Here instead the memory
channel shows additivity whilst improving the transmission rate.

To conclude, for our lossy bosonic memory channel model with a continuous alphabet, and for the particular procedure of encoding (entangled states) and decoding (heterodyne measurement) adopted here, we can assert that  entanglement can be used to enhance the mutual information, in close analogy to what happens for some class of discrete alphabet channels (see, for example, \cite{macc} and \cite{bm}). 

Notice that the presented model well describes the physical situation where all the channel uses are equally correlated each other. This takes place when the block size is such that all the block experiences 
the same memory effect while no correlations among different blocks arise.
However, the model can be easily extend to study a variety of other situations.

\acknowledgments
While completing this work we become aware of a related paper on this subject \cite{cerf}.


\begin{thebibliography}{0}

\bibitem{gall}
Gallager R. G., {\it Information Theory and Reliable Communication}
Wiley, New York (1968).

\bibitem{nich}
Nielsen M. A. and Chuang I. L.
{\it Quantum Computation and Quantum Information}
Cambridge University Press, Cambridge, 
(2003). 

\bibitem{macc} 
Macchiavello C. and Palma G. M., 
Phys. Rev. A {\bf 65} (2002) 050301(R); 
Macchiavello C. Palma G. M. and Virmani S., 
Phys. Rev. A {\bf 69} (2004) 010303(R).

\bibitem{altri}
Hamada M., J. Math. Phys. {\bf 43} (2002) 4382;
Daffer S., W{\'{o}}dkiewicz K. and McIver J. K.,
Phys. Rev. A {\bf 67} (2003) 062312;
Hayashi M. and Nagaoka H.,
IEEE Trans. Inform. Theory {\bf 49} (2003) 1753.

\bibitem{bm} 
Bowen G. and Mancini S.,
Phys. Rev. A {\bf 69} (2004) 012306; 
Bowen G., Devetak I. and Mancini S.,
arXiv: quant-ph/0312216. 

\bibitem{bdb}
Ball J., Dragan A. and Banaszek K., 
Phys. Rev. A {\bf 69} (2004) 042324;
Banaszek K., et al.,
Phys. Rev. Lett. {\bf 92} (2004) 257901.

\bibitem{bp01}
Braunstein S. L. and Pati A. K.
{\it Quantum Information Theory with Continuous Variables},
Kluwer, Dodrecht (2001).

\bibitem{benn}
Bennett C. H. and Shor P. W.,
IEEE Trans.Inf. Theory {\bf 44} (1998) 2724.
Holevo A. S., arXiv:quant-ph/9809023

 \bibitem{giov} 
 Giovannetti V., Guha S., Lloyd S., Maccone L., Shapiro J. H. and Yuen H. P.
 Phys. Rev. Lett. {\bf 92} (2004) 027902.

\bibitem{kr01} 
 Bennett C. H., DiVincenzo D. P. and Smolin J. A.,
Phys. Rev. Lett. {\bf 78} (1997) 3217;
King C. and Ruskai M. B.,
IEEE Trans. Inf. Theory {\bf 47} (2001) 192.

\bibitem{giovman}
Giovannetti V. and Mancini S., 
arXiv: quant-ph/0410176


\bibitem{walls94} 
Walls D. F. and Milburn G. J.
{\it Quantum Optics} Springer-Verlag, Berlin
(1994).

\bibitem{cerf}
Cerf N., Clavareau J., Macchiavello C. and Roland J.,
arXiv:quant-ph/0412089.

\end{thebibliography}
\end{document}